\documentclass[a4paper,11pt]{article}
\usepackage[english]{babel}
\usepackage[all]{xy}
\usepackage[T1]{fontenc}
\usepackage{textcomp}
\usepackage{epsfig}
\usepackage{latexsym}
\usepackage{graphicx}
\usepackage{color}
\usepackage{amsfonts,amssymb,bm}

\newtheorem{proposition}{Proposition}
\newtheorem{theorem}{Theorem}

\newcommand{\beq}{\begin{equation}}
\newcommand{\eeq}{\end{equation}}
\newcommand{\beqa}{\begin{eqnarray}}
\newcommand{\eeqa}{\end{eqnarray}}
\newcommand{\ov}{\overline}

\newcommand{\sign}{\,{\rm sign}\,}

\newcounter{llista}
\newcounter{llista1}

\begin{document}

\title{Flat deformation theorem and symmetries in spacetime.}
\author{Josep Llosa$^1$, Jaume Carot$^2$\\
\small $^1$ Departament de Física Fonamental, Universitat de
Barcelona, Spain\\ \small $^2$ Departament de Física, Universitat de
les Illes Balears, Spain}

\maketitle

\begin{abstract}
The \emph{flat deformation theorem} states that given a
semi-Riemannian analytic metric $g$ on a manifold,  locally there
always exists a two-form $F$, a scalar function $c$, and an
arbitrarily prescribed scalar constraint depending on the point $x$
of the manifold and on $F$ and $c$, say $\Psi (c, F, x)=0$, such
that the \emph{deformed metric} $\eta = cg -\epsilon F^2$ is
semi-Riemannian and flat.   In this paper we first  show that the
above result implies that every (Lorentzian  analytic) metric $g$
may be written in the \emph{extended Kerr-Schild form}, namely
$\eta_{ab} := a \,g_{ab} - 2 b\,k_{(a} l_{b)}$ where $\eta$ is flat
and $k_a,\, l_a$ are two  null covectors such that $k_a l^a= -1$;
next we show how the symmetries of $g$ are connected to those of
$\eta$, more precisely; we show that if the original metric $g$
admits a Conformal Killing vector (including Killing vectors and
homotheties), then the deformation may be carried out in a way such
that the flat deformed metric $\eta$ `inherits' that symmetry.
\end{abstract}

\section{Introduction \label{S1}}

It has been recently proved \cite{LlosaSoler05} that, given a semi-Riemannian
analytic metric $g_{ab}$ on a manifold ${\cal M}$, there exists a
2-form $F_{ab}$ and a scalar function $c$ such that:
\begin{enumerate}
\item An arbitrary scalar constraint $\Psi(c,F_{ab},x)=0$, $x\in{\cal M}$, is fulfilled and
\item The so-called \emph{`deformed metric'}
\begin{equation}\label{df.1}
 \eta_{ab} = c g_{ab} - \epsilon F^2_{ab}\,\quad{\rm where} \quad \epsilon=\pm1\, \quad {\rm and}\quad F^2_{ab}:=F_{ac}g^{cd}F_{db}
\end{equation}
is semi-Riemannian and flat
\end{enumerate}
This result was called {\em Flat Deformation Theorem\,}. For the
purposes of the present paper, we shall only consider the
four-dimensional Lorentzian case.

The proof of the above theorem was based on the existence of
solutions for a partial differential system that is derived from the
condition that $\eta_{ab}$ is flat. As a consequence of the
arbitrariness in the choice of the Cauchy hypersurface and Cauchy
data, the deformation (\ref{df.1}) leading to a flat $\eta_{ab}$ is
by no means unique. Furthermore, as the Cauchy-Kovalewski theorem is
a cornerstone in the proof, the validity of the theorem is limited
to the analytic category.

The purpose of the present paper is to deal with the question of how
the symmetries of the metric $g_{ab}$ are reflected upon the
deformed metric $\eta_{ab}$, more precisely: assuming that $g_{ab}$
admits a Killing vector field $X^a$, we ask whether it is  possible
to choose $F_{ab}$ and $c$ in (\ref{df.1}) such that $X^a$ is also a
Killing vector field for $\eta_{ab}$. We shall prove that the answer
is in the affirmative {in the case of non-null Killing vectors} and
that the symmetry is thus somehow \emph{`inherited'} along the
deformation.

The paper is structured as follows: section \ref{S2} contains some
algebraic developments on the consequences of the deformation law
(\ref{df.1}) for a 4-dimensional spacetime which  will allow us to
state it in a number of alternative ways, thus illustrating
different features of the deformation law. In section \ref{S3} we
present the formalism and prove some intermediate
results\footnote{This formalism was developed in a number of
references, notably \cite{Geroch71} and \cite{Beig00} which will be
used in section \ref{S4}. We present it here in a way well suited to
our purposes.} in order to demonstrate the theorem alluded to in the
previous paragraph. It is worth noticing that in order to prove it,
the problem is reformulated on the 3-dimensional quotient manifold
(see section \ref{SS4.1}), so that a dimensional reduction occurs.
Section \ref{S5} contains a generalization of the above result to
the case of (non-null) Conformal Killing Vectors. Finally, in
section \ref{S6}, we present some examples which we believe may be
of interest due to their physical relevance. We put some technical
developments in the appendices in order to make the paper more
readable. {Also for this reason, we do not insist at every
intermediate step on the local character of the results presented
here, but the reader should bear this in mind.}

\section{Algebraic consequences of the deformation law  \label{S2}}
Consider now the 2-form $F_{ab}$ whose existence is granted by the
deformation theorem \cite{LlosaSoler05}; there are two
possibilities, either it is
\begin{description}
\item[(a) singular (or null),]  then, a tetrad  $\{x_a,\,y_a,\,k_a,\,l_a\}$ exists such that $\,
g_{ab}=x_a x_b + y_a y_b - 2 k_{(a} l_{b)}$ and
\begin{equation}\label{df.2}
 F_{ab} =  2 k_{[a} x_{b]} \qquad {\rm and\; then} \qquad F^2_{ab} = - k_a k_b
\end{equation}
or else it is
\item[(b) non-singular (or non-null),] in which case a tetrad
such as the one above exists in terms of which $F_{ab}$ reads
\begin{equation}\label{df.3}
 F_{ab} =  -2 B\,x_{[a} y_{b]} + 2 E\, k_{[a} l_{b]}\quad {\rm and\; then} \quad
 F^2_{ab} = - B^2\,\left(x_a x_b + y_a y_b\right) - 2 E^2 \,k_{(a} l_{b)}
\end{equation}
where $E$ and $B$ are functions related to the algebraic invariants
of $F^a_{\;b}:= g^{ac} F_{cb}$. If either $B$ or $E$ is zero, the resulting 2-form
is  timelike or spacelike respectively. If neither of them vanishes,
the 2-form is said to be non-simple.
\end{description}

In the singular case, the deformation law (\ref{df.1}) reads $
\eta_{ab} = c\,g _{ab} + \epsilon\, k_a k_b $ or, equivalently,
\begin{equation}\label{df.4}
g_{ab} = \frac1{c}\,\eta _{ab} -\frac{\epsilon}{c}\, k_a k_b
\end{equation}
with $k_a k^a =0$ and $\eta _{ab}$ flat. That is, $g _{ab}$ is a
conformal Kerr-Schild metric \cite{Rafels00}. The singular case is
therefore non-generic and encompasses a rather restricted class of
metrics.

In the non-singular case, from equations (\ref{df.1}) and
(\ref{df.3}) we have that:
\begin{equation}\label{df.5}
\eta_{ab} = a\, g_{ab} + b\, S_{ab}
\end{equation}
with $a=c+\epsilon B^2\,$, $b = -\epsilon\,(B^2 + E^2)\,$ and $\,
S_{ab} = - 2 \,k_{(a} l_{b)}$. As it was shown in
\cite{LlosaSoler05}, this is the generic case in the sense that the
flat deformation (\ref{df.5}) can always be achieved for any
analytic semi-Riemannian metric.

Notice that the arbitrary scalar constraint  $\Psi(c,F_{ab},x)=0$
has no consequences on the factors $a$ and $b$ in (\ref{df.5}).
Indeed, including (\ref{df.3}) the scalar constraint may be written
as $f(c,E,B) =0$ or, equivalently, as a relation $\tilde f(c,a,b)=0$
which, at most, can be used to determine $c=c(a,b)$ to choose one
amongst the many 2-forms $F_{ab}$ compatible with (\ref{df.5}).

We have hitherto proved that:
\begin{proposition}\label{pr.1}
Let $g_{ab}$ be  a  Lorentzian analytic metric on a spacetime ${\cal
M}$. Locally there exist two scalars, $\,a\,$ and $\,b$, and two
null vectors, $k_a$ and $l_a$, such that $k_a l^a = -1$ and the
metric
\begin{equation} \label{df.7}
 \eta_{ab} := a \,g_{ab} - 2 b\,k_{(a} l_{b)}
\end{equation}
is Lorentzian and flat.
\end{proposition}
The above expression vaguely reminds a conformal Kerr-Schild
transformation, but in the present case two non-parallel null
vectors, $k_a$ and $l_a$, occur. We shall henceforth call this expression
{\em extended Kerr-Schild form} and proposition \ref{pr.1} can be restated as:
\begin{quotation}
Any Lorentzian analytic metric can be written in extended Kerr-Schild form.
\end{quotation}

An equivalent statement is
\begin{proposition}\label{pr.2}
Let $g_{ab}$ be  a  Lorentzian analytic metric on a spacetime ${\cal
M}$. Locally there exist two scalars, $\,a\,$ and $\,b$, and a
hyperbolic 2-plane $S_{ab}$  such that the metric
\begin{equation} \label{df.8}
 \eta_{ab} := a \,g_{ab} + b\,S_{ab}
\end{equation}
is Lorentzian and flat.
\end{proposition}

Notice that $S^a_{\;b}$ is a 2-dimensional projector:
\begin{equation} \label{df.9}
 S^a_{\;d} S^d_{\;b} = S^a_{\;b} \,, \qquad  S^a_{\;a} = 2
\end{equation}
which projects vectors onto the hyperbolic plane spanned by
$\{k^a,\,l^a\}$. If we now denote $H_{ab}:= g_{ab}-S_{ab}$, i.e. the
complementary projector, then:
\begin{equation} \label{df.10}
H^a_{\;d} H^d_{\;b} = H^a_{\;b} \,, \qquad  H^a_{\;a} = 2 \,, \qquad
{\rm and} \qquad S^a_{\;d} H^d_{\;b} = H^a_{\;d} S^d_{\;b}  =0
\end{equation}

$H_{ab}$ is then the elliptic 2-plane spanned by any two spacelike
vectors orthogonal to $S_{ab}$, in particular  $x^a, y^a$, the
spacelike vectors in the chosen tetrad, i.e. $H_{ab}= 2x_{(a}y_{b)}$,
and it is then possible to write the deformation
(\ref{df.1}) in a way similar to that given by (\ref{df.8}) but in
terms of the (elliptic) projector $H_{ab}$ instead of the $S_{ab}$,
namely:
\begin{equation} \label{df.8b}
 \eta_{ab} := \bar a \,g_{ab} +\bar b\,H_{ab}
\end{equation}
where $\bar a$ and $\bar b$ are scalars.  

From the comments and developments above and taking (\ref{df.8}) into account, we can  write
\begin{equation} \label{df.11}
 g_{ab}:= H_{ab}+ S_{ab} \qquad {\rm and} \qquad \eta_{ab}:= (a+b)\,S_{ab}+ a\,H_{ab} \,,
\end{equation}
that is, the almost-product structure \cite{Kobayashi} defined by
$S_{ab}$ is compatible with both metrics, $g_{ab}$ and $\eta_{ab}$,
and therefore we can state

\begin{proposition}\label{pr.3}
Let $g_{ab}$ be a  Lorentzian analytic metric on a spacetime ${\cal
M}$.  Locally it exists a Lorentzian flat metric $\eta_{ab}$ that
shares with $g_{ab}$ an almost-product structure.
\end{proposition}

\section{Spacetimes admitting a (non-null) Killing vector \label{S3}}
In this section we are going to set up and develop the formalism and
basic results which will be later used in order to prove the result
stated in the introduction; namely: that if the metric admits an
isometry, it is always  possible to preserve it in the flat deformed
metric.

Let ${\cal M}$ be a spacetime with an arbitrary  metric
$\eta_{ab}$\footnote{Note: $\eta_{ab}$ does not designate the flat
metric at this point. We use this notation here for later
convenience.} admitting a Killing vector $X^a$. Let
$\xi_a:=\eta_{ab} X^b$ and $l:=\xi_a X^a$. Assume that the Killing
is non-null, that is: $l \neq 0$,  and denote by ${\cal S}$ the set
of all orbits of $X^a$, which we assume to be a 3-manifold (the
quotient manifold)\footnote{It can be shown that locally this is
always the case if fixed points of $X^a$ are excluded.}.

{We shall designate by $\pi$ the canonical projection
$\pi: \mathcal{M}  \longrightarrow  \mathcal{S} $
where $\pi (x) = O_x$ is the orbit through the point $x \in
\mathcal{M}$ of the 1-parameter group generated by $X^a$. }

The projector:
\begin{equation} \label{df.12}
h^a_b :=\delta^a_b -\frac1{l} X^a \xi_b
\end{equation}
projects vectors in $T{\cal M}$ onto vectors that are transverse
(orthogonal) to $X^a$.  There is a bijection \cite{Geroch71} between
tensor fields $T^{\prime\,a\ldots}_{b\ldots}$ on ${\cal S}$ and the
tensor fields $T^{a\ldots}_{b\ldots}$ on ${\cal M}$ that fulfill:
\begin{equation} \label{df.13}
 X^b T^{a\ldots}_{b\ldots} = 0 \,, \qquad \xi_a T^{a\ldots}_{b\ldots}=0
 \qquad {\rm and} \qquad \mathcal{L}_XT^{a\ldots}_{b\ldots}=0
\end{equation}
that is, those which are transverse to $X^a$ and $\xi_a$ and Lie
invariant along $X^a$. Following Geroch \cite{Geroch71}
\guillemotleft  While it is useful conceptually to have the
three-dimensional manifold $\mathcal{S}$, it plays no further
logical role in the formalism. We shall hereafter drop the primes:
we shall continue to speak of tensor fields being \emph{on}
$\mathcal{S}$, merely as  a shorthand way of saying that the field
(formally, on $\mathcal{M}$) satisfies (\ref{df.13})\guillemotright

As $l\neq 0$ the projected metric
\begin{equation} \label{df.14}
 h_{ab} :=\eta_{ab} -\frac1{l} \xi_a \xi_b
\end{equation}
induces a semi-Riemannian metric on the quotient manifold ${\cal
S}$, the so-called `quotient metric'. Its signature is $+1\,+1\,-\sign(l)$.
We shall designate by $\displaystyle{h^{ab} :=\eta^{ab} -\frac1{l} X^a X^b}$ the inverse
quotient metric, that is: $h^{ab} h_{bc} = h^a_ c$.

\subsection{The Killing equation \label{SS3.1}}
From $\mathcal{L}_X\eta_{ab}=0$ it follows that $\nabla_a\xi_b$ is
skew-symmetric, that is: $\nabla_b\xi_{a} + \nabla_a \xi_{b}  = 0$
where  $\nabla$  stands for the covariant derivative associated
to $\eta$. 

We also have that $\mathcal{L}_X\xi_a=0$ and $X^a l_a=0$, where $l_a
:= \nabla_a l$. Since $X^a$ is non-null, $ \nabla_a \xi_{b}$ can be
decomposed as:
\begin{equation} \label{df.16}
 2 \nabla_a \xi_b  := 2 \,f_{[a}\xi_{b]} + \Theta_{ab} \qquad {\rm with} \qquad
 f:=\log|l| \qquad {\rm and} \qquad \Theta_{ab}X^b =0
\end{equation}
$\Theta_{ab} = -\Theta_{ba}$ is related with the vorticity of the Killing flow.
We shall use the above form for the Killing equation in the sequel.


\subsection{The Levi-Civita connection on ${\cal S}$ \label{SS3.2}}
Let $T^{a\ldots}_{b\ldots} $ be a tensor field on ${\cal S}$ and
define:
\begin{equation} \label{df.17}
D_c T^{a\ldots}_{b\ldots} := h^a_m h^n_b h^k_c \nabla_k
T^{m\ldots}_{n\ldots}
\end{equation}
Clearly, it is a tensor field on ${\cal S}$, since
$T^{a\ldots}_{b\ldots}$ and $h^{a}_b$ both satisfy (\ref{df.13}),
and, since $X^a$ is a KV, the Lie derivative with respect to it
commutes with  $\nabla$; further it can be easily proved that $D_a$
is a linear connection: indeed, it is linear, it satisfies the
Leibniz rule and for any scalar function $f$ on ${\cal S}$, $D_a f$
is the gradient of $f$. Moreover, it can be also shown that it is
torsion-free and that $D_c h_{ab}=0$ (this last result holds
trivially); therefore, $D$ is the Levi-Civita connection on ${\cal
S}$ (see \cite{Geroch71}).

Let now  $v^a$, $w^b$ be two vector fields on ${\cal S}$, then
taking into account (\ref{df.13}), (\ref{df.16}) and (\ref{df.17})
one easily gets
\begin{equation} \label{df.17a}
D_v w^{a} = \nabla_v w^{a} + \frac1{2l}\, X^a \Theta_{bc} v^b w^c
\end{equation}
where $D_v w^a := v^b D_b w^a$.
Notice the formal similarity between this formula and Gauss equation
for hypersurfaces, even though ${\cal S}$ is not a submanifold and
we have the skew-symmetric $\Theta_{bc}$ instead of the second
fundamental form.

\subsection{The Riemann tensor on ${\cal S}$ \label{SS3.3}}
Consider next a vector field $v^a$ on $\mathcal{S}$ endowed with the
quotient metric $h_{ab}$ and its associated Levi-Civita connection
$D_a$ as defined above in (\ref{df.17}). We aim at calculating the
Riemann tensor $\mathcal{R}^c_{\;dab}$ for this connection.

{From the Ricci identities, $[D_a,D_b] v^c = v^d
\mathcal{R}^c_{\;dab}$, we have that
$${\cal R}_{abcd} = R^\perp_{abcd} +\frac1{2l}\,\left(\Theta_{ab}\Theta_{cd} + \Theta_{[ac}\Theta_{b]d} \right)\,,$$
where $R^\perp_{abcd}:= h^m_a h^n_b h^p_c h^q_d R_{mnpq}$. Using the
identity  $\Theta_{ab}\Theta_{cd} + \Theta_{ac}\Theta_{db} +
\Theta_{ad}\Theta_{bc} = 0$ that follows from the fact that ${\rm
dim}\,{\cal S}=3$, we then arrive at
\begin{equation} \label{df.18}
{\cal R}_{abcd} = R^\perp_{abcd} +\frac3{4l}\,\Theta_{ab}\Theta_{cd}
\end{equation}

The remaining components of $R_{abcd}$ follow from the second order
Killing equation \cite{KSMH2},  $\nabla_a\nabla_b\xi_c = R_{dabc}
X^d := R_{Xabc}$ which, taking into account (\ref{df.16}), leads to:
\begin{eqnarray}
R^\perp_{Xabc} & = & \frac12\,D_a\Theta_{bc} + \frac12\,f_{[b} \Theta_{ac]} \label{df.19}  \\[1ex]
R_{XaXc} & = & -\frac{1}{2}\,D_a l_c  -\frac14\, \Theta_a^{\;\, b} \Theta_{bc} + \frac1{4 l}\, l_a l_c  \label{df.20}
\end{eqnarray}
We have thus shown that the entire Riemann tensor on ${\cal M}$
may be expressed in terms of the kinematic invariants of $\xi_a$ and
the Riemann tensor on ${\cal S}$ associated to the Levi-Civita
connection $D_a$ of the projected (quotient) metric $h_{ab}$. }

\subsection{Lift of a metric from ${\cal S}$ to ${\cal M}$ \label{SS3.4}}
We have hitherto shown how a semi-Riemannian metric can be projected
from ${\cal M}$ to ${\cal S}$. We shall now consider the converse
case. As before, let $X^a$ be a vector field on ${\cal M}$ and let
${\cal S}$ be the set of its orbits, which we take to be a manifold
according to the reasoning at the beginning of the present section.
Further, let $\pi: {\cal M} \rightarrow {\cal S}$ be the canonical
projection.

Let now $h_{ab}$ be a semi-Riemannian metric on ${\cal S}$ {having
constant signature $(+\,+\,\sigma)$, $\sigma=\pm 1$}. We shall
denote by the same symbol the pulled back metric on ${\cal M}$,
i.e.: $\pi^* h_{ab} = h_{ab}$, which is degenerate because
$h_{ab}X^b =0$, moreover, $\mathcal{L}_X h_{ab}=0$. The point now
is: does it exist a metric $\eta_{ab}$ on ${\cal M}$ such that: (a)
admits $X^a$ as a Killing vector and (b) has $h_{ab}$ as the
quotient metric?

If it exists, a relation similar to (\ref{df.14}) must hold, with
$\xi_a :=\eta_{ab}X^b$ and $l=\xi_a X^a$. Hence, the solution is not
unique, because we may choose any covector  $\xi_a$ such that
$\mathcal{L}_X\xi_a =0$ {and that $l:=\xi_a X^a$ has constant
sign\footnote{The sign is to be chosen so that the lifted metric has
the required signature $(+\,+\,+\,-)$}; then taking}
\begin{equation}\label{df.21}
\eta_{ab}:= h_{ab} + \frac1{l}\,\xi_a \xi_b
\end{equation}
as the lifted metric, all the required conditions are satisfied
(namely: $X^a$ is a KV of $\eta_{ab}$ and $h_{ab}$ is its quotient
metric). Then, if no further condition is demanded, equations
(\ref{df.18}), (\ref{df.19}) and (\ref{df.20}) merely relate the
Riemann tensors for both metrics, $\eta_{ab}$ and $h_{ab}$. However,
if we require the lifted metric $\eta_{ab}$ to fulfill some
supplementary condition, e.g. to be flat, then these become
equations on the chosen $\xi_a$ and the given $h_{ab}$, much in the
same way as the Gauss curvature equation and the  Codazzi-Mainardi
equations set up conditions on the way that a submanifold can be
immersed in an ambient space.

{The choice of $\xi_a$ is restricted by the condition
$\mathcal{L}_X\xi_a =0$. Assume that a 1-form $\alpha_a$ on ${\cal
M}$ such that $\alpha_a X^a=1$ and $\mathcal{L}_X\alpha_a =0$  is
given. Then, the  sought $\xi_a$ can be written as $\xi_a =
l(\alpha_a + \mu_a)$, with $l:= \xi_a X^a$ and $\mu_a X^a =0$. It
can be easily proved that:
$$ \mathcal{L}_X\xi_a = 0\qquad \Leftrightarrow \qquad X l = 0 \quad {\rm and} \quad \mathcal{L}_X\mu_a = 0 $$

Hence, given a 1-form $\alpha_a$ on ${\cal M}$ such that $ \alpha_a X^a=1 $ and $\mathcal{L}_X\alpha_a =0$, choosing
$\xi_a $ is equivalent to choosing a function $l\neq 0 $ on ${\cal S}$, a 1-form $\mu_a$ on ${\cal S}$ and taking
$\xi_a = l(\alpha_a + \mu_a)$.

The exterior derivative of this expression yields
\begin{equation} \label{df.23c}
\left( d\xi\right)_{ab} = \frac2{l}\, l_{[a} \xi_{b]} + l \left( d
\mu \right)_{ab}  \qquad {\rm and} \qquad \Theta_{ab} = l \left( d
\mu \right)_{ab} + l \left(d\alpha\right)_{ab}
\end{equation}
where (\ref{df.16}) has been taken into account.

In terms of $l$ and $\mu_a$, taking (\ref{df.23c}) into account, the equations (\ref{df.18}),
(\ref{df.19}) and (\ref{df.20}) read:
\begin{eqnarray} \label{df.25a}
R^\perp_{abcd} &=& {\cal R}_{abcd} - \frac{3l}{4}\,(d\mu)_{ab} (d\mu)_{cd} \\[.5ex]
\label{df.25b}
R^\perp_{Xabc} &=& \frac12\,D_a\left[ l(d\mu)_{bc} \right] + \frac12\,l_{[b}(d\mu)_{ac]} \\
\label{df.25c}
R_{XaXc} &=& -\frac12\, D_a l_c -\frac{l^2}{4}\,(d\mu)_{ad} (d\mu)_{bc} h^{bd}
\end{eqnarray}
that are equations for $l$, $\mu_a$ and $h_{ab}$ to be solved on ${\cal S}$. }

\subsection{Hypersurfaces and Killing vectors \label{SS3.4.5}}
Let $\Sigma$ be a surface in ${\cal S}$, then $\pi^{-1}\Sigma$ is a
hypersurface in ${\cal M}$ and  the Killing vector $X^a$ is tangent
to it. The following diagram is commutative:
\begin{displaymath}
\xymatrix{(\eta,\nabla, R) \quad & \mathcal{M}\ar[r]^\pi & \mathcal{S}  & \quad (h,D,{\cal R})\\
(\eta^\prime,\nabla^\prime, R^\prime) \quad & \pi^{-1}\Sigma \ar[r]^\pi \ar[u]^J & \Sigma \ar[u]_j & \quad (h^\prime,D^\prime,
{\cal R}^\prime) }
\end{displaymath}

where $J$ and $j$ are the respective embeddings.

We respectively denote by $\eta^\prime_{ab}$, $\Phi_{ab}$,
$\nabla^\prime$ and $R^\prime_{abcd}$ the first and second
fundamental forms, the induced connection and the intrinsic
curvature on $\pi^{-1}\Sigma$ as a hypersurface of the Riemannian
manifold $({\cal M},\eta_{ab})$. Similarly, we denote by
$h^\prime_{ab}$, $\phi_{ab}$, $D^\prime$ and ${\cal
R}^\prime_{abcd}$ the corresponding objects on $\Sigma$ regarded as
a hypersurface in $({\cal S}, h_{ab})$.

Let $n^a$ be the unit vector $\eta$-normal to $\pi^{-1}\Sigma$.
Since $X^a$ is tangent to $\pi^{-1}\Sigma$, then $\xi_a n^a=0$.
Furthermore, $\mathcal{L}_X n^a =0$.
Indeed, for any $V^a$ tangent to $\pi^{-1}\Sigma$ we
have that $\mathcal{L}_X V^a$ is also tangent to $\pi^{-1}\Sigma$ and,
using that $X^a$ is a Killing vector field, we easily arrive at $ \eta_{ab}
\mathcal{L}_X n^a V^b= 0$, which implies that $\mathcal{L}_X n^a
\propto n^a$. On the other hand, as $n^a$ is unit, $ \eta_{ab}
\mathcal{L}_X n^a n^b= 0$, whence it follows that  $\mathcal{L}_X
n^a =0$. Therefore, $n^a$ is also a vector in ${\cal S}$ and is the
unit vector $h$-normal to $\Sigma$.

It can be easily proved that the second fundamental forms for $\pi^{-1}\Sigma$ and $\Sigma$ satisfy that:
$ \phi_{ab}=\Phi^\perp_{ab} $.
On the other hand, for any vector field $V^b$ tangent to $\pi^{-1}\Sigma$, we have that
$$\Phi_{ab} X^a V^b = \nabla_X n_b V^b = - \nabla_V \xi_b n^b = -\frac12\,(d\xi)_{ab} V^a n^b $$
where in the second equality we have used that $\mathcal{L}_X V^a
n_a =0$ and that $V^bn_b=0$. The above equation implies, putting $
(d\xi)_{ab} n^b := (d\xi)_{an}  $ and $f_b n^b := f_n$, \beq
\label{df.352} \Phi_{ab} X^a = \frac12\,(d\xi)_{nb} = \frac12\,f_n
\xi_b +\frac12\,\Theta_{nb} \eeq where (\ref{df.16}) has been taken
into account. Therefore, \beq \label{df.353} \Phi_{ab} = \phi_{ab} +
\frac1{2l}\,f_n\,\xi_a \xi_b + \frac1{l}\,\Theta_{n(a}\xi_{b)} \eeq

\section{Flat deformation \label{S4}}
The aim of this section is to prove the main result in this paper, namely,
\begin{theorem} \label{T1}
Let $({\cal M},g_{ab})$ be a spacetime with a metric $g_{ab}$
admitting a non-null Killing vector $X^a$. Locally there exists a
deformation law
 \begin{equation} \label{df.26}
\eta_{ab} = a \,g_{ab} + b \,H_{ab}
 \end{equation}
where $a$ and $b$ are two scalars, $H_{ab} $ is a 2-dimensional projector on a  $g$-elliptic plane
and $\eta_{ab}$ is flat and also admits $X^a$ as a Killing vector.
\end{theorem}

It will be convenient for our purposes to prove the following result
previously:

\begin{proposition}  \label{P8}
Let $X^a$ be a Killing vector for $g_{ab}$ and let $\eta_{ab}$ be
defined by (\ref{df.26}) with $b\neq 0$, then:
 \begin{equation} \label{df.27}
\mathcal{L}_X\eta_{ab} = 0 \qquad \Leftrightarrow \qquad
\mathcal{L}_X a = \mathcal{L}_X b = 0 \qquad {\rm and} \qquad
\mathcal{L}_X H_{ab}=0
 \end{equation}
\end{proposition}

\paragraph{Proof:} As $\mathcal{L}_X g_{ab} = 0$, $\mathcal{L}_X \eta_{ab}=0$ implies that
 \begin{equation} \label{df.28}
\mathcal{L}_X a \,g_{ab}+  \mathcal{L}_X b \,H_{ab} +
b\,\mathcal{L}_X H_{ab}=0
 \end{equation}
Since $H^a_{b}$ is a 2-dimensional projector, $H^{ab}
H_{ab}=H^a_{a}=2$, and taking the Lie derivative we get
$2\,\mathcal{L}_X H_{ab} H^{ab} = 0$. Contraction of (\ref{df.28})
with $g^{ab}$ and $H^{ab}$  leads respectively to
$$ 4 \mathcal{L}_X a + 2 \mathcal{L}_X  b = 0 \qquad {\rm and } \qquad 2 \mathcal{L}_X  a + 2 \mathcal{L}_X  b = 0 $$
which imply: $\mathcal{L}_X a = \mathcal{L}_X  b = 0$. Substituting
back into (\ref{df.28}) and taking into account that $b\neq 0$
yields $\mathcal{L}_X  H_{ab} = 0$. \hfill $\Box$

The proof of theorem \ref{T1} spreads over the present section and
it consists in finding $a$, $b$ and $H_{ab}$ such that:
\begin{list}
{(\roman{llista})}{\usecounter{llista}}
\item $\eta_{ab}=a g_{ab} + b H_{ab}$ is flat and
\item $\mathcal{L}_X a = \mathcal{L}_X b = 0$ and $\mathcal{L}_X H_{ab} =0$.
\end{list}

The number of unknowns is 6, namely: 2 for $a$ and $b$ plus 4 for
$H_{ab}$ (recall the constraints $H^a_c H^c_b = H^a_b$ and $H^a_a
=2$). Then, (i) means that the Riemann tensor for $\eta_{ab}$
vanishes:
\begin{equation} \label{df.30}
  R_{abcd}=0
  \end{equation}
To ensure (ii) we shall solve (\ref{df.30}) on ${\cal S}$ and then pull the solutions back to $\pi^{-1}{\cal S} = {\cal M}$.

We first introduce the decompositions:
\begin{equation} \label{df.31}
g_{ab} = p_{ab} +\frac1{\bar{l}}\,\bar\xi_a \bar\xi_b \qquad {\rm and} \qquad
\eta_{ab} = h_{ab} +\frac1{{l}}\,\xi_a \xi_b
\end{equation}
where $\bar\xi_a :=g_{ab} X^b $ and $\bar l =\bar\xi_b X^b$ are known from the data $g_{ab}$ and $X^b$, whereas
\begin{equation} \label{df.32}
\xi_a:=\eta_{ab} X^b = a \bar\xi_a + b H_{ab} X^b \qquad {\rm and} \qquad
l :=\xi_a X^a \,,
\end{equation}
depend on the unknowns. Notice that $b H_{ab}X^a X^b = l - a \bar l $.

\subsection{The projection of our problem onto the quotient manifold ${\cal S}$ \label{SS4.0}}
We must now replace the unknowns $(a,\,b,\, H_{ab})$, which are
tensor quantities on ${\cal M}$, with others that are tensor
quantities on ${\cal S}$. Consider the covector
$\alpha_a=\bar\xi_a/\bar{l}$. It is obvious that $\alpha_a X^a=1$
and $\mathcal{L}_X \alpha_a =0$;  hence the results in section
\ref{SS3.4} can be applied and we have that $\displaystyle{\mu_a =
\frac1{l}\,\xi_a - \frac1{\ov l}\,\ov\xi_a }$ is a covector in
${\cal S}$, thus we can write
\begin{equation} \label{df.B2}
\xi_a = l m \nu_a + \frac{l}{\ov l}\, \ov\xi_a \,,
\end{equation}
where $\nu_a$ is a $p$-unitary covector on ${\cal S}$ and
$m:=\sqrt{\mu_a \mu_b p^{ab}}$.  Then, on account of (\ref{df.32}),
we have that\footnote{We explicitly exclude the cases $l-a\bar l =
0$ and $b=0$ since they are non-generic. Note that $b=0$ corresponds
to the metric $\eta$ being conformally flat.}:
\begin{equation} \label{df.B3}
H_{ab} X^b = \frac{1}{b}\,\left( l m\,\nu_a + \frac{l-a\ov l}{\ov l}\,\ov\xi_a\right)
\end{equation}

Now, $H^a_{\;b}$ is a 2-dimensional projector and therefore its
eigenvalues are 0 and 1, both with multiplicity 2.  $H^a_{\;b} X^b$
is an eigenvector (not unit), and a second one may be chosen so that
it is $g$-orthogonal to it. We can thus write:
\begin{equation} \label{df.B4}
H_{ab}= \beta_a \beta_b +\omega_a \omega_b
\end{equation}
where $\beta_a$ and $\omega_a$ are $g$-unitary and mutually $g$-orthogonal, and
\begin{equation} \label{df.B5}
\beta_a =\frac{lm}{\sqrt{b(l-a\ov l)}}\,\nu_a + \frac{1}{\ov l}\,\sqrt{\frac{l-a\ov l}{b}}\, \ov\xi_a
\end{equation}

Since $H^a_{\;b}$ is a projector, it follows that $\omega_a X^a =
\omega_a \nu^a = 0$, and as $\beta_a$ is $g$-unitary we also have
that:
\begin{equation} \label{df.B6}
\frac{l^2m^2}{l-a\ov l} = b + a - \frac{l}{\ov l}
\end{equation}

From $\mathcal{L}_X H_{ab}=0$ (Proposition \ref{P8}), its transverse
projection
\begin{equation} \label{df.B7}
\tilde H_{ab} = \frac{(b+a)\ov l - l}{b \ov l}\,\nu_a \nu_b +\omega_a \omega_b
\end{equation}
satisfies also $\mathcal{L}_X \tilde H_{ab}=0$. Hence, $\tilde
H_{ab}$ is a tensor on ${\cal S}$.

The quotient metric $h_{ab}$ is the transverse projection of
$\eta_{ab}$ and, taking (\ref{df.26}), (\ref{df.31}) and (\ref{df.B7}) into
account, we obtain:
\begin{equation} \label{df.B7a}
h_{ab} = a p_{ab} + \left( b + a - \frac{l}{ \ov l}\right)\,\nu_a \nu_b + b \,\omega_a \omega_b
\end{equation}

We have seen so far that the set of unknowns $\{a,\,b,\,H_{ab}\}$
---tensor quantities on ${\cal M}$---  can be assigned the new set
of unknowns $\{a,\,b,\, l,\,\nu_a,\, \omega_b\}$, where $\nu_a$ and
$\omega_b$ are $p$-unitary and mutually $p$-orthogonal covectors on
${\cal S}$, and $a$, $b$ and $l$ are scalar functions on ${\cal S}$.
The inverse correspondence is easily established. It suffices to
take $H_{ab}$ as defined by (\ref{df.B4}) with $\beta$ defined by
(\ref{df.B5})

(Notice that the number of degrees of freedom is still 6 because,
once $\nu_a$ is given, the unit orthogonal covector $\omega_a$ is determined by only giving one angle.)\\[1ex]

Due to the symmetries of the Riemann tensor, $R_{abcd}$, it can be separated as:
\begin{equation} \label{df.33}
R_{abcd} = L_{abcd} +\frac2{l}\,\left( L_{ab[c}\xi_{d]} + L_{cd[a}\xi_{b]} \right) + \frac4{l^2}\,\xi_{[b} L_{a][c}\xi_{d]}
\end{equation}
where $L_{abcd}$, $L_{abc}$ and $L_{ac}$ are transverse to $X^b$ and have the following  symmetries:
\begin{list}
{(\alph{llista})}{\usecounter{llista}}
\item $L_{abcd}$ has the same symmetries as a Riemann tensor in 3 dimensions,
\item $\quad L_{abc}= - L_{bac}$, $\quad L_{abc} + L_{bca} + L_{cab} = 0 \;$ and $\;L_{ab} = L_{ba}$
\end{list}

Notice that:
\begin{equation} \label{df.34}
L_{abcd} = R^\perp_{abcd}\,, \qquad L_{abc} = R^\perp_{abcX} \qquad {\rm and} \qquad
L_{ac} = R_{XaXc}
\end{equation}
and are given by (\ref{df.18}), (\ref{df.19}) and (\ref{df.20}).
Then equations (\ref{df.30}) ---flatness of $\eta_{ab}$--- are
equivalent to:
\begin{equation} \label{df.35}
L_{abcd} = 0\,, \qquad L_{abc} = 0 \qquad {\rm and} \qquad  L_{ac} = 0
\end{equation}

By taking the exterior differential of $ m \nu_a$ and taking (\ref{df.B2}) and (\ref{df.23c}) into account, we have that
\begin{equation} \label{df.B3a}
2 D_{[a} \left(m \nu_{b]}\right) =\frac1{l}\,\Theta_{ab} - \frac1{\ov l}\,\ov\Theta_{ab}
\end{equation}
with $m$ given by (\ref{df.B6}). Including now (\ref{df.18}), (\ref{df.19}), (\ref{df.20}), (\ref{df.B7a}) and (\ref{df.B3a}), the
equations (\ref{df.35}) result in second order partial differential
equations relating $a$, $b$, $l$, $\nu_a$ and $\omega_b$, i.e. tensor quantities
on ${\cal S}$.

\subsection{The constraints and the reduced system \label{SS4.1}}
Equations (\ref{df.35}) constitute a system of 20 independent
equations for only 6 independent unknowns. To handle this
overdetermination we shall take 6 equations among them as a {\em
reduced partial differential system} \cite{FriedRen}, that we shall
solve by giving Cauchy data on a non-characteristic surface $\Sigma$
\cite{John}. The remaining 14 equations are to be considered as {\em
constraints} to be fulfilled by the Cauchy data on $\Sigma$. It must
be then proved that any given solution of the reduced PDS fulfilling
the constraints on $\Sigma$ also fulfills them on a neigbourhood of
$\Sigma$.

Given a surface $\Sigma\subset  {\cal S}$, we choose Gaussian
$p$-normal coordinates $(x^1,x^2,x^3)$ on a neigbourhood ${\cal U}
\subset {\cal S}$ of $\Sigma$:
\begin{equation} \label{df.B9}
 x^1 =0 \qquad {\rm on}\quad \Sigma \,, \qquad p_{11}= s = \pm 1 \qquad {\rm and}\qquad p_{1j}=0\,, \quad j=2,3
\end{equation}
{The sign $s$ depends on the sign of $\ov l$: if $\ov l<0$, then
$s=+1$, while for $\ov l>0$, $s$ can take both values $\pm 1$. For
the sake of simplicity, here we shall choose $\Sigma$ so that
$s=-\sign(\ov l)$ and then $p_{ij}$ has signature $(+\,+)$.}

In these coordinates, we choose (indices $a$, $b$, $c$, \ldots run from 1 to 3 and $i$, $j$, \ldots run from 2 to 3)
\begin{equation} \label{df.36}
L_{11}= 0\,, \qquad L_{1j1}=0 \,, \qquad L_{1i1j} =0
\end{equation}
as the {\em reduced partial differential system} and
\begin{equation} \label{df.37}
L_{aj}=0\,, \qquad  L_{bijk}= 0\,, \qquad  L_{jcd}=0
\end{equation}
as the {\em constraints}. (Notice that $L_{1jk}=0$ is included in
the above equalities because, as a consequence of the first Bianchi
identity, $L_{1jk}=-L_{jk1}-L_{k1j}$.)

In Appendix A we prove that, if $a$, $b$, $H_{ab}$ is  an analytic
solution of the reduced PDS (\ref{df.36}) fulfilling the constraints
(\ref{df.37}) on $\Sigma$, then the constraints are also fulfilled
in an open neigbourhood of $\Sigma$.

\subsection{The reduced PDS \label{SS4.2}}
We shall now write equations (\ref{df.36}) in terms of  the unknowns
$\{a,\,b,\,l,\, \nu_a,\, \omega_a\}$. We shall only make explicit
the principal parts, i.e. those terms involving second order partial
derivatives with respect to the coordinate $x^1$. In what follows a
``dot'' will stand for  $\partial_1$, whereas $\cong$ will mean
``equal apart from non-principal terms''.
\begin{list}
{(\alph{llista})}{\usecounter{llista}}
\item From (\ref{df.34}) and (\ref{df.20}), and taking into account that $l\neq 0$, we have that ${L}_{11} =0$ leads to
  \beq \label{df.49}
  \ddot l  \cong 0\,.
  \eeq
\item From (\ref{df.34}) and (\ref{df.19}), including (\ref{df.B3a}), we obtain
$\, L_{abc} \cong - D_c \left[l\,D_{[a}(m \nu_{b]})\right]\,$.
Therefore,  $L_{1j1}=0$ amounts to
\beq  \label{df.53}
\ddot m \nu_j + m \ddot \nu_j  \cong 0 \,, \qquad j=2,3
\eeq
with $m$ given by (\ref{df.B6}).

\item From (\ref{df.34}) and (\ref{df.18}) we have that
the third of the equations (\ref{df.36}) $L_{1i1j}=0$ leads to
\beq \label{df.56}
   \ddot h_{ij} \cong 0
\eeq which, using (\ref{df.B7}), (\ref{df.B7a}), (\ref{df.49}) and
(\ref{df.53}), becomes
\beq  \label{df.B11}
\ddot a \,\left( p_{ij} + [\ov l ( b+a) - l] \nu_i \nu_j \right) +  \ddot b\,\omega_i\omega_j +
b\left[\ddot \omega_i \omega_j + \omega_i \ddot\omega_j \right]
\cong 0 \,, \qquad \qquad i,j=2,3
\eeq

The characteristic determinant for the reduced partial differential
system constituted by the six equations (\ref{df.49}), (\ref{df.53})
and (\ref{df.B11}) is (see Appendix B for details):
\begin{eqnarray*}
\Delta & := & 2 b \omega_1^2 \nu_1 \tau_1\,p\, \left[1 - s\omega_1^2 + [\ov l ( b+a) - l] \nu_1^2\right] \, \frac{l-a\ov l}{l^2}\,
\nonumber \\
 & & \left[\left(b+a-\frac{l}{\ov l}\right)\,(1-s\nu_1^2) - s\omega_1^2 \,\left(a-\frac{l}{\ov l}\right)\right]
\end{eqnarray*}

\end{list}

\subsection{Geometrical meaning of the constraints \label{SS4.4}}
It remains to be shown that Cauchy data fulfilling the constraints
(\ref{df.37}) on the Cauchy surface $\Sigma$ do exist. Consider
$\pi^{-1}\Sigma$, which is  a hypersurface in ${\cal M}$, and take
coordinates $(x^1, \ldots x^4)$ adapted to both $X^a$ and
$\pi^{-1}\Sigma$, i.e. $X^a=\delta^a_4$ and $x^1=0$ on
$\pi^{-1}\Sigma$.

Let $\ov P^a_b$ and $P^a_b$ be the projectors:
\beq \label{df.441}
\ov P^a_b := \delta^a_b - \frac1{g^{11}}\,g^{1a}\,\delta^1_b \qquad \qquad {\rm and}  \qquad \qquad
 P^a_b := \delta^a_b - \frac1{\eta^{11}}\,\eta^{1a}\,\delta^1_b
\eeq {They both project vectors in $T{\cal M}$ onto the hyperplane
$T(\pi^{-1}\Sigma)$ and, while $\ov P^a_b$ projects parallelly to
$g^{1a}$, $ P^a_b$ does it parallelly to $\eta^{1a}$. It is obvious
that $\ov P^1_b = P^1_b =0$, hence \beq \label{df.442} \ov P^a_b
P^b_c = P^a_c \qquad \qquad {\rm and}  \qquad \qquad  P^a_b \ov
P^b_c = \ov P^a_c \eeq which implies that, when restricted to the
hyperplane $T(\pi^{-1}\Sigma)$, both projectors, $\ov P^a_b$ and
$P^a_b$, yield the identity.}

It is easy to see that the constraints (\ref{df.37}) amount to
\begin{center}
\centerline{$R_{abcd} = 0 \qquad$ whenever at most one of the indices is 1}
\end{center}
that is, $R_{abcd} \ov P^b_e \ov P^c_f \ov P^d_g = 0$  which,
including (\ref{df.442}) is equivalent to \beq \label{df.443}
R_{abcd} P^b_e P^c_f P^d_g = 0 \eeq Then, if $n^a $ is the unit
vector $\eta$-normal to $\pi^{-1}\Sigma$, (\ref{df.443}) is
equivalent to \beq \label{df.444} R_{abcd}^{\rm tang} = 0 \qquad
\qquad {\rm and} \qquad \qquad R_{nbcd}^{\rm tang} = 0 \eeq where
``tang'' denotes components tangential to  $\pi^{-1}\Sigma$ and
$R_{nbcd}:=R_{abcd} n^a$.

$\pi^{-1}\Sigma$ can be seen both as a hypersurface of the
Riemannian  manifold $({\cal M},\eta_{ab})$ and as a hypersurface of
$({\cal M},g_{ab})$. We shall denote $\eta_{ab}^\prime$ and
$g_{ab}^\prime$ the respective first fundamental forms. The two
normal vectors are respectively:
\begin{equation}  \label{df.445a}
 n^a= \frac1{\sqrt{|\eta^{11}|}}\,\eta^{1a}\,, \qquad n_a = \frac1{\sqrt{|\eta^{11}|}}\,\delta^1_a \qquad{\rm and} \qquad
\ov n^a= \frac1{\sqrt{|g^{11}|}}\,g^{1a} \,,  \qquad  \ov n_a = \frac1{\sqrt{|g^{11}|}}\,\delta^1_a
\end{equation}
and the second fundamental forms are:
$$ \Phi_{ab} = P^c_a \nabla_c n_b \qquad \qquad {\rm and} \qquad \qquad \ov \Phi_{ab} = \ov P^c_a \ov\nabla_c \ov n_b $$

The Gauss curvature equation for $\pi^{-1}\Sigma$ as a submanifold
of $({\cal M},\eta_{ab})$ reads \cite{Hicks}: \beq \label{df.445}
R_{abcd}^{\rm tang} = R_{abcd}^\prime + 2\,\Phi_{a[d} \Phi_{c]b}
\eeq and the Codazzi-Mainardi equation is \beq \label{df.446}
R_{nbcd}^{\rm tang} = 2\,\nabla^\prime_{[d} \Phi_{c]b} \eeq where
$\nabla^\prime$ and $R_{abcd}^\prime$ are respectively the induced
connection and the intrinsic curvature.

The constraints (\ref{df.444}) are therefore equivalent to
$$ R_{abcd}^\prime + 2\,\Phi_{a[d} \Phi_{c]b} = 0 \qquad \qquad {\rm and} \qquad \qquad \nabla^\prime_{[d} \Phi_{c]b}=0  $$
a particular solution of which is \beq \label{df.447} \Phi_{ab} = 0
\qquad \qquad {\rm and} \qquad \qquad  R_{abcd}^\prime =0 \eeq

\paragraph{The normal derivatives of the unknowns.} The first of the
equations (\ref{df.447}) determines the first order normal
derivatives of  the unknowns on the Cauchy hypersurface $\Sigma$.
Indeed, from (\ref{df.353}) and $\Phi_{ab}=0$ we have that:
\begin{equation}  \label{df.448}
 f_n =0 \,, \qquad \Theta_{nb} = 0 \,, \qquad \phi_{ab}=0
\end{equation}
Furthermore, as $\Theta_{ab}$ is skewsymmetric and $\phi_{ab} n^b =
0$, it is obvious  that $\phi_{ab} = 0$ and $ \Theta_{na} =0 $ are equivalent to
$$ \phi_{ij} = 0 \qquad {\rm and} \qquad \Theta_{nj} = 0\,, \qquad i,j = 2,3 $$
Notice that the remaining equations, namely  $\phi_{a4} = 0$ and $
\Theta_{n4} =0 $, are identically satisfied because $\phi_{ab}$ and
$\Theta_{ab}$ are tensors on ${\cal S}$ and in these coordinates
$X^a= \delta^a_4$.

Including then (\ref{df.353}), (\ref{df.32}) and (\ref{df.B3a}),
equations (\ref{df.448}) are equivalent to:
\begin{equation}  \label{df.449}
 n^b D_b l  =0 \,, \qquad 2 l \, \overline{ D}_{[b} \left(m \nu_{j]}\right)\, n^b + \frac{l}{\overline{l}} \,\overline\Theta_{nj} = 0
 \,, \qquad D_i n_j=0
\end{equation}
and, using (\ref{df.445a}), we have that
$$ D_a n_b = \sqrt{\frac{|g^{11}|}{|\eta^{11}|}}\,
\left(\overline\phi_{ab} + \frac12\,\overline{n}_b D_a\log \left[
\frac{|g^{11}|}{|\eta^{11}|}\right] - b^c_{ab} \overline n_c
\right) $$ where $ b^c_{ab}$ is the difference tensor for the
connections $D$ and $\overline D$.

In Gaussian $p$-normal coordinates, taking into account
(\ref{df.B7a}) and writing explicitly the principal terms only,
(\ref{df.449}) becomes:
\begin{equation}  \label{df.450}
h^{11} \dot l \cong 0\,, \qquad h^{11} \left( \dot m \nu_j  + m \dot \nu_j \right) \cong 0 \,, \qquad h^{11} \dot h_{ij} \cong 0
\end{equation}
The similitude of these equations with (\ref{df.49}), (\ref{df.53})
and (\ref{df.56}) is apparent and the characteristic determinant is
$(h^{11})^6 \Delta$. Hence, provided that the Cauchy data on $\Sigma$ are
chosen so that $\Delta \neq 0$ and $h^{11}\neq 0$, the constraints $\Phi_{ab} = 0$
permit to obtain the first order normal derivatives of the unknowns,
namely $\dot a$, $\dot b$, $\dot l$, $\dot \nu_b$ and $\dot \omega_c$ on $\Sigma$,
in terms of the values of $a$, $b$, $l$, $\nu_b$ and $\omega_c$ on $\Sigma$.

\paragraph{The unknowns on the Cauchy surface $\Sigma$.}
The second of the equations in (\ref{df.447}) is a condition on the
values of the unknowns on $\Sigma$. The isometry group $G$ generated
by $X^a$ acts also on $\pi^{-1}\Sigma$ and $\pi^{-1}\Sigma/G
=\Sigma$. Hence, relations similar to (\ref{df.18}-\ref{df.20})
hold:
\begin{eqnarray} \label{df.a1}
R^{\prime\perp}_{abcd}  & = & {\cal R}^\prime_{abcd} - \frac1{2l}\,\left(\Theta^{\prime}_{ab}\Theta^{\prime}_{cd} + \Theta^{\prime}_{[ac}\Theta^{\prime}_{b]d} \right)= 0  \\[1ex]
R^{\prime\perp}_{Xabc} & = & \frac12\,D^{\prime}_a\Theta^{\prime}_{bc} + \frac12\,f^{\prime}_{[b} \Theta^{\prime}_{ac]} = 0 \label{df.a2}  \\[1ex]
R^{\prime}_{XaXc} & = & -\frac{1}{2}\,D^{\prime}_a l^{\prime}_c -\frac14\, \Theta_a^{\prime\;\;\, b} \Theta^{\prime}_{bc} = 0 \label{df.a3}
\end{eqnarray}
with $R^\prime:=J^\ast R\,$, ${\cal R}^\prime:=j^\ast {\cal R}\,$, $\Theta^\prime := j^\ast \Theta\,$, $l^\prime = j^\ast l$.

As $\Sigma$ has only two dimensions,   $\Theta^\prime_{ac}
\Theta^{\prime\,bc}= \theta^{\prime\,2} h^{\prime\,b}_{\;a}$, where
$2 \theta^{\prime\,2} = \Theta^\prime_{bc} \Theta^{\prime\,bc}$.
Hence, equation (\ref{df.a2}) is equivalent to $\Theta^{\prime\,bc}
R^{\prime\perp}_{Xabc} =0$ which, after a little algebra yields
$D^\prime_a \theta^{\prime\,2} + f^\prime_a \theta^{\prime\,2} =0\,$
and, since $f^\prime = \log|l^\prime|$, we have that:
\begin{equation} \label{df.a4}
 \theta^{\prime\,2} l^\prime = {\rm constant}
\end{equation}
In two dimensions, the Riemann tensor has only one independent component:
$\, {\cal R}^\prime_{abcd} = {\cal R}^\prime \,h^\prime_{a[c}h^\prime_{d]b} \,$,
therefore (\ref{df.a1}) and (\ref{df.a3}) are respectively equivalent to
\begin{equation} \label{df.a4z}
{\cal R}^\prime = \frac{3\theta^{\prime\,2}}{2 l^\prime} \qquad {\rm and} \qquad
 D^\prime_a D^\prime_c l^\prime = \frac12\, \theta^{\prime\,2} \,h^\prime_{ac}
\end{equation}
The integrability conditions for this equation imply that $\theta^\prime = 0$. Indeed, as
$$D^\prime_b D^\prime_a D^\prime_c l^\prime - D^\prime_a D^\prime_ bD^\prime_c l^\prime = -{\cal R}^{\prime\,d}_{\;\;cba} D^\prime_d l^\prime \,,$$
we have that $ D^\prime_{[b}\theta^{\prime\,2} h^\prime_{a]c} = -
{\cal R}^\prime D^\prime_{[b}l^\prime h^\prime_{a]c}\, ,$ where the
fact that we are in 2 dimensions has been used to simplify the
Riemann tensor. Taking now into account the first equation
(\ref{df.a4z}) we obtain $D^\prime_{b}\theta^{\prime\,2}
-\frac{3\theta^{\prime\,2}}{2 l^\prime} D^\prime_b l^\prime$, or
$\theta^{\prime\,2}/l^{\prime\,3}=$constant or
$\theta^{\prime\,2}l^{\prime\,-3}=$constant. This, together with
(\ref{df.a4}) implies $l^\prime=$constant which substituted in
(\ref{df.a4z}) leads to $\theta^\prime=0$.

Therefore, equations (\ref{df.a1}-\ref{df.a3}) are equivalent to:
\begin{equation} \label{df.a8}
{\cal R}^\prime = 0 \,, \qquad \qquad \theta^\prime = 0 \qquad {\rm and} \qquad D^\prime_a D^\prime_c l^\prime = 0
\end{equation}

The Gaussian $p$-normal coordinates introduced in section
\ref{SS4.1}, equation (\ref{df.B9}), are specially well suited to
our problem. In these coordinates vectors that are tangent to
$\Sigma$ are characterized by $v^1=0$ and the restriction to
$\Sigma$ of any covariant tensor on ${\cal S}$, $T_{ab\ldots}\,$,
$\,a,b, \ldots = 1,2,3\,$, merely consists in keeping the components
$T_{ij\ldots}\,$, $\,i,j, \ldots = 2,3\,$. Thus, $h^\prime_{ij} :=
(j^\ast h)_{ij} = h_{ij}\,$, $\nu^\prime_i :=(j^\ast \nu)_{i} = \nu_i\,$,
$\overline\Theta^\prime_{ij} :=(j^\ast \overline\Theta)_{ij} = \overline\Theta_{ij}\,$,
$m^\prime :=m \circ j = m\,$ and so on.

Now, including this and the second equation (\ref{df.a8}),
the restriction to $\Sigma$ of equation(\ref{df.B3a}) is
\begin{equation} \label{df.a10}
 2 D^\prime_{[i} \left( m \nu_{j]}\right) = - \frac1{\ov l}\,\ov\Theta_{ij} \,, \qquad i,j = 2,3
\end{equation}
and, as all differential forms in $\Lambda^2 \Sigma$ are closed,
this equation is locally integrable and yields $m \nu_j\,$, $j=2,3$.

Moreover, $l^\prime = $constant is a solution of the third  equation
(\ref{df.a8}) and therefore we shall take $l = $constant on
$\Sigma$.

As $\Sigma$ has only two dimensions, ${\cal R}^\prime = 2
\epsilon^{\prime\,ij}(h^\prime) \epsilon^{\prime\,kl}(h^\prime)
{\cal R}^\prime_{ijkl}$,  where $\epsilon^{\prime\,ij}(h^\prime)$ is
the volume tensor on $\Sigma$ for the metric $h^\prime_{kl}$. In two
dimensions the volume tensors $\epsilon^{\prime\,ij}(h^\prime)$ and
$\epsilon^{\prime\,ij}(p^\prime)$ are proportional to each other and
therefore ${\cal R}^\prime=0$ is equivalent to
$\epsilon^{\prime\,ij}(p^\prime)
\epsilon^{\prime\,kl}(p^\prime){\cal R}^\prime_{ijkl} = 0$ , or
\begin{equation} \label{df.a11}
p^{ik} p^{jl} {\cal R}^\prime_{ijkl} = 0
\end{equation}
This is a condition on $h^\prime_{ij}$ which depends on the unknowns $a,\, b,\,l,\,\nu_a,\,  \omega_b$, $a,b = 1,\,2,\,3$\,.

From the third equation (\ref{df.a8}) we know that $l=$constant on $\Sigma$.
Then, by solving equation (\ref{df.a10}) we obtain $m \nu_j$, $j=2,\,3$, on $\Sigma$.
We then choose $\omega_i$, $i=2,\,3$, on $\Sigma$ which, together with the orthogonality conditions
$$ \omega_a \omega_b p^{ab}=\nu_a \nu_b p^{ab}=1 \qquad {\rm and} \qquad \nu_a \omega_b p^{ab}=0 \,,$$
permit to obtain $\omega_b$, $\nu_a$, $a,b=1,2,3$ and $m$. Finally,
substituting this in (\ref{df.B6}), we can obtain $b=b(a)$ and
therefore condition (\ref{df.a11}) yields a partial differential
equation for $a$, whose principal part is
\begin{equation} \label{df.912}
\left(\nu^j \nu^k -[1 + p^{il}\nu_i \nu_l]\,p^{jk} \right)\, \partial_{jk} a \cong 0 \qquad {\rm where} \qquad \nu^j:=p^{jk}\nu_k
\end{equation}
The characteristic form is:
$$ \chi(z_l) =  (z_l\nu^l)^2 - [1 + p^{il}\nu_i \nu_l] \,p^{jk}z_j z_k $$
and the existence of non-characteristic lines for equation (\ref{df.a11}) is obvious.

\subsection{Summary of the proof}
So far, we have analyzed the existence of a solution to the problem
stated in section \ref{S1}. Let us now synthesize a way to find such
a solution:
\begin{list}
{(\alph{llista1})}{\usecounter{llista1}}
\item From the given $X^a$ and $g_{ab}$, obtain $\overline l$, $\overline\xi_a$,  $\overline\Theta_{ab}$ and the quotient metric $p_{ab}$;
\item Choose a Cauchy surface $\Sigma \stackrel{j}{\rightarrow} {\cal S}$ and a chart
of Gaussian $p$-normal coordinates for $\Sigma$, $(x^1, x^2, x^3)$;
\item Choose $m \nu_i$, $i=2,3$, on $\Sigma$ as a solution of $\;2 \overline l \partial_{[i}\left( m\nu_{j]}\right)= -\overline\Theta_{ij}$;
\item Take $l = $constant on $\Sigma$;
\item Then choose $\omega_i$, $i=2,3$, such that inequality $\Delta \neq 0$ is fulfilled and,
including the  orthonormality condition, the definition (\ref{df.32}) and the obtained value for
$m \nu_j$, derive $\omega_1$, $\nu_1$ and $m$ on $\Sigma$;
\item With the relation (\ref{df.B6}) obtain $b=b(a)$ and
\item Substitute the above in (\ref{df.a11}) and solve it to obtain $a$ on $\Sigma$.
\end{list}
With this we have $a,\; b,\; l,\; \nu_c, \omega_d$ on $\Sigma$. Then
\begin{list}
{(\alph{llista1})}{\usecounter{llista1}\addtocounter{llista1}{6}}
\item Solve (\ref{df.449}) to derive $\dot a,\; \dot b,\;\dot l,\;\dot\nu_c, \dot\omega_d$ on $\Sigma$; and
\item With these Cauchy data, solve the reduced partial differential system (\ref{df.36});
then use (\ref{df.B2}) to have $\xi_a$, (\ref{df.B7a}) to have $h_{ab}$ and (\ref{df.31}) to have $ \eta_{ab}$.
\end{list}

\section{Generalization to Conformal Killing Vectors}\label{S5}

The main result in this paper, stated in theorem \ref{T1}, can be
extended almost immediately to the case  of Conformal Killing
Vectors (CKV for short), as a consequence of the so called
Defrise-Carter's theorem (see for instance \cite{Hall2004}); which
states, roughly speaking, that given a (non-conformally flat) metric
$g$ admitting an $r$-dimensional Lie algebra of CKVs,  $C_r$ , there
exists a function $\Omega$, such that $C_r$ becomes a Lie algebra of
Killing vectors for the conformally related metric $\tilde g= \Omega^2 g$.

Thus, we can state:
\begin{theorem} \label{T2}
Let $(\mathcal{M},g)$ be a spacetime such that the metric $g_{ab}$
admits a non-null CKV $X^a$. Locally, there exists a deformation law
as the one given by (\ref{df.26}) such that $X^a$ is a KV for the
flat metric $\eta_{ab}$.
\end{theorem}

\paragraph{Proof:} Since $X^a$ is a CKV of the metric $g_{ab}$,
there exists a conformal factor $\Omega^2$ such that $\tilde
g_{ab}:= \Omega^2 g_{ab}$ has $X^a$ as a KV \cite{Hall2004}. By
theorem \ref{T1}, it then follows that  a flat, deformed metric
$\eta_{ab}$ exists, $$ \eta_{ab} = \tilde a\tilde g_{ab} + bH_{ab}$$
for which $X^a$ is a KV, defining next $ a:=\Omega^2 \tilde a$ and
taking into account the above expression for $\eta_{ab}$ as well as
the relation between the metrics $g$ and $\tilde g$, it readily
follows that $X^a$ is a KV of the flat metric
$$ \eta_{ab} = a g_{ab} + bH_{ab}.$$
\hfill $\square$

\section{Examples}\label{S6}
We next present some physically significant examples. We have chosen
families of well characterized spacetimes and then selected, amongst
all spacetimes in the family, one well-known and physically relevant
particular solution. {For the sake of  convenience, instead of the
deformation law (\ref{df.26}) in theorem \ref{T1}, we shall rather
use the equivalent formula (\ref{df.8}) with the hyperbolic
projector $S_{ab}$.}

\subsection{Class A1 warped spacetimes \label{S6.1}}
For these spacetimes, coordinates $x^a = u, x^k$ with $k=1,2,3$
exist such that the metric takes the following form (see
\cite{CarotdaCosta1993} for definitions and further details),
$$ ds^2 = \epsilon du^2 + f^2(u) h_{ij}(x^k) dx^idx^j, \quad
\epsilon=\pm 1$$ where $f$ is some function of $u$. For $\epsilon =
+1$, $u$ is a spacelike coordinate (class A1 spacelike warped),
whereas for $\epsilon = -1$, $u$ is time (class A1 timelike warped).
In what follows, we shall consider only the latter case and put
$u:=t$, thus, we shall take the line element to be
\begin{equation}\label{6.1.1}
ds^2 = - dt^2 + f^2(t) h_{ij}(x^k) dx^idx^j, \; \;\; i,j,k =
1,\ldots,3.
\end{equation}

Writing now
\begin{equation}\label{6.1.2} ds^2 = \tilde f^2(\tau) d\tilde s^2,
\qquad\mathrm{with}\qquad d\tau = \frac{dt}{f(t)}, \quad \tilde
f(\tau) = f(t(\tau))
\end{equation}
we get, in an obvious notation,
\begin{equation}\label{6.1.3}  d\tilde s^2 = -d\tau^2 + p_{ij}(x^k)
dx^idx^j, \qquad \mathrm{or\; else}\qquad g_{ab} = \tilde f^2(\tau)
\tilde g_{ab}.
\end{equation}

Now, $\partial_\tau$ is a KV of $\tilde g_{ab}$ and a CKV of the
original metric $g_{ab}$; further, it is orthogonally transitive.
hence, $p_{ij}(x^k)$ is a Riemannian metric on the quotient manifold
coordinated by $x^k, \; k=1,2,3$.

Making use of the equivalent to the flat deformation theorem in
three dimensions  for a Riemannian metric (see
\cite{CollLlosaSoler2003}), we can see that a scalar function
$a(x^k)$ and a covariant vector field $\mu_i(x^k)$ exist such that
they fulfill a  previously chosen arbitrary relation, say $\Psi
(a,||\mu||)=0$, where $ ||\mu||^2 = p^{ij}\mu_i\mu_j$, with
$p^{ij}p_{jk}=\delta^i_k$,  and the metric
\begin{equation}\label{6.1.4}
\hat\eta_{ij}= a p_{ij} + \mu_i\mu_j
\end{equation}
is flat. Presently, we choose
$$\Psi (a,||\mu||)=||\mu||^2 + a -1 =0,$$
and it then follows that the 4-dimensional semi-Riemannian metric
$$\eta := -d\tau \otimes d\tau + \hat\eta_{ij} (x^k) dx^i\otimes
dx^j$$ is also flat and admits the KV $\partial_\tau$.

Using now (\ref{6.1.4}) we have that
$$ \eta := -d\tau \otimes d\tau + a p_{ij} dx^i
\otimes dx^j + \mu_i dx^i \otimes \mu_j dx^j,$$ or else,
using the coordinates $x^a = x^1, x^2,x^3, x^4=\tau$, setting $\mu_4
=0$ and making use of (\ref{6.1.2}), it turns out that we can write
\begin{equation}\label{6.1.5}
\eta_{ab} = a \tilde g_{ab} - (1-a)
\delta^4_a\delta^4_b + \mu_a\mu_b = a\tilde g_{ab} + (1-a) S_{ab},
\end{equation}
where
$$ S_{ab} :=-\delta^4_a\delta^4_b + \hat\mu_a \hat\mu_b,
\qquad \hat\mu_a := \frac{1}{||\mu||} \mu_a$$
is a two-dimensional hyperbolic projector (recall that we chose $||\mu||^2 =1-a $), and thus
(\ref{6.1.5}) corresponds the sought for form (\ref{df.8}).

\subsection{Spacetimes with additional symmetries}
In some cases with additional symmetries it is possible to derive an explicit expression for $\mu_i$; this giving for granted that the deformed metric $\eta_{ab}$, the factors $a$ and $b$, and the hyperbolic projector $S_{ab}$ will share the same additional symmetries. (Notice that this is only a conjecture that goes beyond what has been proved so far, although theorem \ref{T1} supports its plausibility.)

As an example, take a static spherically symmetric metric
\begin{equation}\label{schw1}
    g= -f^2(r) dt\otimes dt +p^2(r) dr\otimes dr + r^2\left(d\theta \otimes d\theta + \sin^2\theta  d\phi \otimes  d\phi
    \right).
\end{equation}
which, besides the three KV implementing the spherical symmetry, admits one fourth KV, namely $\partial_t$. The quotient space ${\cal S}$ can be given the structure of a manifold as discussed previously. Consider next the metric $h$ on $\mathcal{S}$,
\begin{equation}\label{schw2}
    h= g + f^2(r) dt\otimes dt = p^2(r) dr\otimes dr + r^2\left(d\theta \otimes d\theta + \sin^2\theta  d\phi \otimes  d\phi
    \right).
\end{equation}
By the theorem in \cite{CollLlosaSoler2003} regarding three-dimensional metrics, a scalar $a$ and a covariant vector
$\mu_i$ exist, which fulfill an arbitrary, previously chosen constraint, that we shall take $\Psi (a,||\mu||) := \|\mu\|^2 - f^{-2}(r) + a= 0$, and are such that the deformed
three-dimensional Riemannian metric
\begin{equation}\label{schw3}
    \hat \eta= a h + \mu \otimes \mu
\end{equation}
is flat. Let us next make a guess at $a$ and $\mu$ and take $a=a(r)$
and $\mu = \mu(r) dr$, we shall have:
$$ ||\mu||^2 = h^{ij} \mu_i\mu_j = p^{-2}(r) \mu^2(r),$$
hence
\begin{equation}\label{schw4}
    \hat \eta= (a+||\mu||^2) \, p^{2}(r) dr\otimes dr + a r^2\left(d\theta \otimes d\theta + \sin^2\theta  d\phi \otimes  d\phi
    \right).
\end{equation}
The spacetime metric $\eta := -dt\otimes dt + \hat\eta $ is also flat, i.e.:
\begin{equation}\label{schw5}
     \eta= -dt\otimes dt + ah + \mu \otimes \mu = a \left( g + f^2(r)\,dt\otimes dt \right) + \mu \otimes \mu -dt\otimes dt,
\end{equation}
which is already in the desired form (\ref{df.8}) with
$ bS := \mu \otimes \mu - \left( f^{-2}(r) - a \right)\,f^2(r)\, dt\otimes dt \,.$

In order to ensure that $S$ is a hyperbolic projector as required, we need $\|\mu\|^2 = f^{-2}(r) - a$ which is fulfilled thanks to the chosen arbitrary constraint  $\Psi (a,||\mu||) = 0$.

Substituting the above back into (\ref{schw4}) we get that
\begin{equation}\label{schw6}
 \hat\eta=  f^{-2}(r) p^{2}(r) dr\otimes dr + a r^2 \left(d\theta \otimes d\theta + \sin^2\theta d\phi \otimes d\phi \right)
\end{equation}
must be flat, and this  determines $a$ up to a constant. Notice that a line element of the form
$$ d\sigma^2 = F^2(r) dr^2 + Y^2(r)d\Omega^2$$
is flat iff
$$ \frac{d Y(r)}{dr} = \pm F(r) \,,$$
hence, choosing the plus sign for convenience and since $Y^2 = a r^2$ and $F(r) = p(r)/f(r)$, we finally get
\begin{equation}\label{6.2}
\sqrt{a} = \frac1{r}\,\left(\int^r \frac{p(r^\prime)}{f(r^\prime)}\,dr^\prime + K \right)\,,
\qquad K = \mathrm{constant.}\qquad {\rm and } \qquad \mu = p(r)\,\sqrt{f^{-2}(r) - a}
\end{equation}

Two interesting particular cases are the following:
\begin{description}
\item[Friedmann-Robertson-Walker spacetimes]{These are particular instances of the the ones just discussed,
namely: class A1 timelike warped. As it is well known, the metric
may be written as
\begin{equation}\label{6.2.1}
    ds^2 =-dt^2 + \frac{R^2(t)}{1+\frac{k}{4}r^2} \left( dr^2 + r^2d\Omega^2
    \right), \qquad  d\Omega^2 = d\theta^2 +\sin^2\theta d\phi^2.
\end{equation}
Proceeding as in the general case in section \ref{S6.1}, we can write $ds^2 =R^2(t) d\tilde s^2$, where
\begin{equation}\label{6.2.2}
     d\tilde s^2:= -d\tau^2 + \left(1+\frac{k}{4}r^2\right)^{-1} \left( dr^2 + r^2d\Omega^2
    \right) \;\;\mathrm{and}\;\; d\tau := \frac{dt}{R(t)},
\end{equation}
with $\partial_\tau$ being a KV of the metric $\tilde g$ (of line element $d\tilde s^2$) and a CKV of $g$ (line element $d s^2$).

The metric $\tilde{g}$ is a particular case of (\ref{schw1}) with
$$ f(r) := 1 \qquad {\rm and} \qquad p(r) := \left(1 +\frac{k}{4}\, r^2\right)^{-1/2}  $$
which substituted in (\ref{6.2}) yield
$$\mu = \frac{1 - a}{\sqrt{1 +k r^2/4}} $$
and

$$ \sqrt{a} = \frac1{r}\,\left( K + \int^r dr'\,\left[1+\frac{k}{4}\, r'^2\right]^{-1/2} \right)  $$}
\item[Schwarzschild solution]{Consider next the well-known Schwarzschild solution written in the form
\begin{equation}\label{schw1a}
    g= -\left(1-\frac{r_s}{r} \right) dt\otimes dt +\left(1-\frac{r_s}{r} \right)^{-1} dr\otimes
    dr + r^2\left(d\theta \otimes d\theta + \sin^2\theta  d\phi \otimes  d\phi
    \right).
\end{equation}
which is a particular case of (\ref{schw1}) with
$$ f(r) := \sqrt{1-\frac{r_s}{r}} \qquad {\rm and} \qquad p(r) := 1/f(r)  $$
which substituted in (\ref{6.2}) yield
$$\mu = \sqrt{\left(1-r_s/r\right) -a \left(1-r_s/r\right)^2} $$
and
$$\sqrt{a} = 1 + \frac{r_s}{r} \left[K +\ln\left(\frac{r}{r_s}-1 \right)\right]\,,
\qquad K = \mathrm{constant.}$$}
\end{description}

\section*{Acknowledgements}
The authors are grateful to B. Coll, A. Molina and J.M. Pozo for
interesting comments and suggestions upon reading a previous version
of the manuscript. J.C.  acknowledges financial support from the
Spanish Ministerio de Educación through grant No. FPA-2007-60220.
Partial financial support from the Govern de les Illes Balears is
also acknowledged. J.Ll. acknowledges financial support from
Ministerio de Educación through grant No. FIS2007-63034 and from the
Generalitat de Catalunya, 2001SGR-00061 (DURSI). Both authors are
also grateful to an anonymous referee for his comments and
suggestions, which have greatly contributed to making the paper more
precise and readable.

\section*{Appendix A: }
We here prove that the constraints (\ref{df.37}) propagate out of
$\Sigma$.  Assume that $a$, $b$ and $H_{ab}$ is a solution of the
reduced PDS (\ref{df.36}) for a set of Cauchy data fulfilling the
constraints (\ref{df.37}) on the Cauchy surface $\Sigma$. We must
prove that these constraints also hold on a neighbourhood of
$\Sigma$.

Given $a$, $b$ and $H_{ab}$, consider the metric $\eta_{ab} = a
g_{ab} + b H_{ab}$. Let $\nabla$ and  $R_{abcd}$ respectively denote
the Levi-Civita connection and the Riemann tensor for $\eta_{ab}$.
By the second Bianchi identity we have that:
\begin{equation} \label{df.A1}
\sum_{\{cde\}} \nabla_e R_{abcd} := \nabla_e R_{abcd} + \nabla_c R_{abde} + \nabla_d R_{abec} \equiv 0
\end{equation}
Including (\ref{df.33}), the different projections of this equation are
\begin{list}
{(\alph{llista})}{\usecounter{llista}}
\item the projection on $ X^b$ is:
  \beq \label{df.A12a}
  \sum_{\{cde\}} \,\left(  D_e  L_{cda} - \frac12\, f_e L_{cda} +
  \frac1{l}\, L_{ac} \Theta_{ed} -\frac12\, L_{abcd}\Theta_e^{\;b}\right) \equiv 0
\eeq
\item the totally transverse projection yields:
  \beq \label{df.A12b}
\sum_{\{cde\}} \left(D_e  L_{abcd} +  \frac1{l}\, L_{abc} \Theta_{ed}
+  \frac1{l}\, L_{cd[a}\Theta_{eb]} \right) \equiv 0
\eeq
\item and the projection on $X^e$  is:
\beq \label{df.A12c}
\nabla_X R_{abcd} + 2 \nabla_{[c} R_{abd]X} - 2 R_{ab[de} \nabla_{c]} X^e = 0
\eeq
which is transverse to $X$ for the indices  $c$ and $d$.

As $X^a$ is a  Killing vector, $\mathcal{L}_X R_{abcd} = 0$, and the
above equation becomes \beq \label{df.A12}
 \nabla_{[c} R_{abd]X} -  R_{e[bcd} \nabla_{a]} X^e = 0
 \eeq
which, projected on $X^b$ and including (\ref{df.33}), yields
\beq \label{df.A13}
D_{[c}L_{ad]} - \frac12\,L_{ab[d} \Theta_{c]}^{\;\;b} -\frac12\, f_{[c} L_{ad]} - \frac14\, L_{cdb}\Theta_{a}^{\;\;b} - \frac{l}{4}\, L_{bacd} f^b = 0
\eeq
On its turn, the totally transverse component of (\ref{df.A12}) is:
\beq \label{df.A14}
D_{[c} L_{abd]} - \frac12\,L_{e[bcd} \Theta_{a]}^{\;\;e} + \frac12\,L_{cd[b} f_{a]} + \frac1{l}\,\Theta_{[c[b} L_{a]d]} = 0
\eeq
\end{list}

In Gaussian normal coordinates equations (\ref{df.A12a}), (\ref{df.A12b}), (\ref{df.A13}) and (\ref{df.A14}) respectively read:
\beq  \label{df.A16}
\left. \begin{array}{lcl}
 \partial_1 L_{jka}+ 2 \partial_{[j} L_{k]1a} = {\rm lin\,} \,, &\qquad \qquad &  \partial_1 L_{abjk}+ 2 \partial_{[j} L_{abk]1} = {\rm lin\,} \\
\partial_1 L_{aj} - \partial_j L_{a1} = {\rm lin\,}\,,&\qquad \qquad & \partial_1 \partial_1 L_{abj} - \partial_{j} L_{ab1} = {\rm lin\,}
\end{array}
\right\} \eeq where $j=2,3$ and $a,b,\ldots=1,2,3$, and ``lin''
denotes ``linear terms not containing partial  derivatives''. (We
have only kept those equations governing the propagation outwards of
$\Sigma$, i. e. those containing partial derivatives with respect to
$x^1$.)

As the metric $\eta_{ab}$ is a solution of the reduced PDS
(\ref{df.36}), we have that $L_{11}=0\,$, $L_{1j1}=0 \,$ and
$L_{1i1j}=0$.  Equations (\ref{df.A16}) thus yield the following
linear partial differential system to be fulfilled by the
constraints (\ref{df.37}):
\begin{eqnarray*} 
\partial_1 L_{jkl}  = {\rm lin\,} + 2 \partial_{[j} L_{k]l1} \,, & \qquad &
\partial_1 L_{jk1}  = {\rm lin\,} \qquad \\
\partial_1 L_{lijk}  = {\rm lin\,} + 2 \partial_{[j} L_{lik]1} \,, & \qquad &
\partial_1 L_{1ijk}  = {\rm lin\,} \qquad
{\rm and} \\
\partial_1 L_{1j}  = {\rm lin\,} \,, & \qquad &
\partial_1 L_{ij}  = {\rm lin\,} + \partial_j L_{i1}
\end{eqnarray*}
which is already in the normal form for the Cauchy-Kowalevski
theorem \cite{John}. As the chosen solution $a$, $b$ and $H_{ab}$ of
(\ref{df.36}) is assumed to be analytic, the coefficients are
analytic. Then, for the Cauchy data $L_{aj}=0\,$, $L_{bijk}= 0$ and
$L_{jcd}=0$ on $\Sigma$, the solution is unique in the analytic
category and, by linearity, $L_{aj}=0\,$, $L_{bijk}= 0$ and
$L_{jcd}=0$ on an open neighbourhood of $\Sigma$.

\section*{Appendix B: The characteristic determinant}
The reduced PDS is constituted by the six equations (\ref{df.49}), (\ref{df.53}) and (\ref{df.B11}):
\begin{eqnarray}
& & \ddot l  \cong  0\,  \label{df.138a}\\
& & \ddot m \nu_j + m \ddot \nu_j  \cong 0 \,, \qquad j=2,3  \label{df.138b}\\
& & \ddot a \,\left( p_{ij} + [\ov l ( b+a) - l] \nu_i \nu_j \right) +  \ddot b\,\omega_i\omega_j + b\left[\ddot \omega_i \omega_j + \omega_i \ddot\omega_j \right] \cong 0 \,, \qquad \qquad i,j=2,3 \label{df.138c}
\end{eqnarray}
where
$$ \ddot m =\frac{m}2\,\left(\frac{\ddot b + \ddot a }{b + a - l/\ov{l}} + \frac{\ddot a }{a - l/\ov{l}}  \right) $$
as it easily follows from (\ref{df.B6}) and (\ref{df.138a}).

The surface $\Sigma$ is non-characteristic if the PDS  can be solved
for the second partial derivatives of the unknowns, namely  $\ddot a
$, $\ddot b $, $\ddot l$, $\ddot \nu_a$ and $\ddot\omega_b$ on
$\Sigma$, where a ``double dot'' stands for $\partial_1^2$. Notice
that due to the constraints of $p$-unitarity and $p$-orthogonality,
in $\ddot \nu_a$ and $\ddot\omega_b$ there are only three
independent unknowns. In order to handle them more appropriately we
shall consider the $p$-orthonormal triad of spatial covectors
$$ \omega_a\,, \, \nu_a\,,\, \tau_a \qquad {\rm where} \qquad \tau_a:= \ov\epsilon_{abc}\omega^b\nu^c,  $$
where $\ov\epsilon_{abc}:=\ov\epsilon_{abcd} X^d/\ov l$ is the $p$-volume tensor on ${\cal S}$.

We then have that:
$$ \dot\omega_a = \Omega_3\nu_a - \Omega_2\tau_a\,, \qquad
\dot\nu_a = -\Omega_3\omega_a + \Omega_1\tau_a\,, \qquad
\dot\tau_a = \Omega_2\omega_a - \Omega_1\nu_a  $$
and, deriving again and keeping only principal terms:
\begin{equation} \label{df.C1}
\ddot\omega_a = \dot\Omega_3\nu_a - \dot\Omega_2\tau_a\,, \qquad
\ddot\nu_a = -\dot\Omega_3\omega_a + \dot\Omega_1\tau_a\,, \qquad
\ddot\tau_a = \dot\Omega_2\omega_a - \dot\Omega_1\nu_a
\end{equation}
which introduced in (\ref{df.138b}) and (\ref{df.138c}) yields
\begin{eqnarray}
& & \ddot m \nu_j - m \omega_j \dot\Omega_3 + m \tau_j \dot\Omega_1 \cong 0 \,, \qquad j=2,3  \label{df.139b}\\
& & \ddot a \,\left( p_{ij} + [\ov l ( b+a) - l] \nu_i \nu_j \right) +  \ddot b\,\omega_i\omega_j +
2 b \nu_{(i} \omega_{j)}\dot\Omega_3  - 2 b \tau_{(i} \omega_{j)}\dot\Omega_2 \cong 0 \,, \qquad  i,j=2,3 \label{df.139c}
\end{eqnarray}

This last expression (\ref{df.139c}) contains three independent
equations, which amount to the contractions with $p^{ij}$, $\omega^i
\omega^j - p^{ij} \omega^l \omega^l$ and $\nu^i \nu^j - p^{ij} \nu^l
\nu^l$. They  read, respectively:
\begin{equation} \label{df.140}
\left. \begin{array}{l}
\left(2 + [\ov l ( b+a) - l] \nu_l \nu^l \right)\,\ddot a  +  \omega^l\omega_l\, \ddot b+ 2 b \nu^j \omega_j\, \dot\Omega_3 - 2 b\tau^j \omega_j\, \dot\Omega_2 \cong 0     \\
\left[-\omega^l\omega_l + [\ov l ( b+a) - l] \left( (\nu_l \omega^l)^2 - \nu_l \nu^l\, \omega^j \omega_j \right)\right] \, \ddot a \cong 0  \\
- \nu_l \nu^l\,\ddot a + \left( (\nu_l \omega^l)^2 - \nu_l \nu^l\, \omega^j \omega_j \right)\,\ddot b
- 2 b\left( \nu_l \omega^l\,\nu_j \tau^j - \nu_l \nu^l\, \tau^j \omega_j \right)\,\dot\Omega_2 \cong 0
\end{array} \right\}
\end{equation}
On its turn, the  expression (\ref{df.139b}) consists of two
independent equations.  They are equivalent to the wedge products
with $\tau_i$ and $\nu_i$, namely
\begin{equation} \label{df.141}
\left. \begin{array}{l}
- m \,(\nu\wedge\omega)\,\dot\Omega_3 +  m \,(\nu\wedge\tau)\,\dot\Omega_1 \cong 0 \\
\displaystyle{(\tau\wedge\nu)\,\frac{m}2\,\left(\frac{\ddot b + \ddot a }{b + a - l/\ov{l}} + \frac{\ddot a }{a - l/\ov{l}}  \right)  - m \,(\tau\wedge\omega)\,\dot\Omega_3 \cong 0   }
\end{array} \right\}
\end{equation}
where (\ref{df.B6}) has been used and $(\nu\wedge\omega):=\nu_2\omega_3-\nu_3\omega_2$ and so on.

Some simplification is gained taking into account that
$\{\omega_a,\,\nu_b,\,\tau_c\}$ is a $p$-orthonormal triad and, in
the Gaussian $p$-normal coordinates of section \ref{SS4.1}, we have
that:
$$ \omega\wedge\nu = s \tau_1\sqrt{p}\,, \qquad  \nu\wedge\tau = s \omega_1\sqrt{p}\,, \qquad  \tau\wedge\omega = s \nu_1\sqrt{p}$$
where $p:=\det(p_{ij}) $, and
$$ (\omega^l \nu_l)^2 - \omega^l \omega_l\nu^j \nu_j = -\frac1p\,(\nu\wedge\omega)^2 = - \nu_1^2\,, \qquad
\omega^l \nu_l\nu^j \tau_j - \omega^l \tau_l\nu^j \nu_j = \omega_1 \tau_1 $$
Furthermore,
$$\omega^l \omega_l = 1 - s\omega_1^2\,, \qquad \nu^j \tau_j = - s \nu_1 \tau_1 \,, \qquad \omega^j \tau_j = - s \omega_1 \tau_1 $$
Substituting this into (\ref{df.138a}), (\ref{df.140}) and (\ref{df.141}), we obtain
\begin{equation} \label{df.140a}
\left. \begin{array}{l}
\left(2 + [\ov l ( b+a) - l] (1-s\nu_1^2) \right)\,\ddot a  +  (1-s\omega_1^2)\, \ddot b - 2 s b \nu_1 \omega_1\, \dot\Omega_3 + 2 s b\tau_1 \omega_1\, \dot\Omega_2 \cong 0     \\
\left[-1+s\omega_1^2 - [\ov l ( b+a) - l] \nu_1^2\right] \, \ddot a \cong 0  \\
- (1-s\nu_1^2)\,\ddot a  - \nu_1^2\,\ddot b - 2 b \omega_1\tau_1\,\dot\Omega_2 \cong 0  \\
 m \, s \tau_1\sqrt{p} \,\dot\Omega_3 +  m \,s \omega_1\sqrt{p} \,\dot\Omega_1 \cong 0 \\
\displaystyle{- \frac{m}2\,s \omega_1\sqrt{p} \,\left(\frac{\ddot b + \ddot a }{b + a - l/\ov{l}} + \frac{\ddot a }{a - l/\ov{l}}  \right) - m \,
s \nu_1\sqrt{p}\,\dot\Omega_3 \cong 0   }
\end{array} \right\}
\end{equation}

The reduced PDS (\ref{df.138a})-(\ref{df.138c}) can be solved for
all the second partial derivatives of the unknowns,  namely  $\ddot
a $, $\ddot b $, $\ddot l$, $\ddot \nu_a$ and $\ddot\omega_b$, if,
and only if, the system (\ref{df.140a}) can be solved for the six
unknowns $\ddot a $, $\ddot b $, $\ddot l$, $\dot\Omega_1$,
$\dot\Omega_2$ and $\dot\Omega_3$; that is if, and only if, it has a
non-null determinant, $\Delta\neq 0$, where
\begin{eqnarray}
\Delta & := & 2 b \omega_1^2 \nu_1 \tau_1\,p\, \left[1 - s\omega_1^2 + [\ov l ( b+a) - l] \nu_1^2\right] \, \frac{l-a\ov l}{l^2}\,
\nonumber \\
 & & \left[\left(b+a-\frac{l}{\ov l}\right)\,(1-s\nu_1^2) - s\omega_1^2 \,\left(a-\frac{l}{\ov l}\right)\right]
\label{df.200}
\end{eqnarray}
which stands for the characteristic determinant of the partial
differential system  (\ref{df.138a}-\ref{df.138c}).


\end{document}